\documentclass[twocolumn,final,aps,prb,showpacs,superscriptaddress,amsmath,amssymb,amsfonts,floatfix]{revtex4-1}
\usepackage{epsfig}
\usepackage[dvipsnames,usenames]{color}
\usepackage{color}
\usepackage{amsmath}
\usepackage{comment}
\usepackage{array}
\usepackage{multirow}
\usepackage{tabularx}
\usepackage{float}
\usepackage[colorlinks=true,bookmarks=true, pdfborder={0 0 0},linkcolor=blue,urlcolor=blue,    breaklinks=true,bookmarksnumbered=true]{hyperref}

\emergencystretch=\maxdimen
\begin{document}
\title{Anisotropic multi-orbital Hubbard model simulated with impurity approximation}
\author{Yan Peng}%\,\orcidlink{0000-0002-9500-202X}}
%\email{jiangmi@suda.edu.cn}
\affiliation{School of Physical Science and Technology, Soochow University, Suzhou 215006, China}

\author{Mi Jiang}%\,\orcidlink{0000-0002-9500-202X}}
%\email{jiangmi@suda.edu.cn}
\affiliation{School of Physical Science and Technology, Soochow University, Suzhou 215006, China}
\affiliation{Jiangsu Key Laboratory of Frontier Material Physics and Devices, Soochow University, Suzhou 215006, China}

\begin{abstract}
Motivated by the recent experimental findings on the orbital ordering of cuprate SC, we have investigated the multi-orbital Hubbard model in the framework of Cu impurity approximation embedded in the O lattice by incorporating the 3d$^{8}$ multiplet structure coupled to a full O-2p band. 
Our systematic investigation on the impact of anisotropy of various parameters reveal rich phenomena in terms of the ground state (GS) weight asymmetry between $\hat{x}$ and $\hat{y}$ directions.
The numerical evidence demonstrate that the GS weight of Zhang-Rice singlet (ZRS) can be affected by the asymmetry of these parameters to distinct extent. Although the experimentally motivated asymmetric charge transfer energy only induces tiny weight difference, the asymmetric $d$-$p$ hybridization can result in considerable change of the weight. Besides, the nearest-neighbor $V_{pd}$ has much stronger impact than the local $U_{pp}$, which stems from the nature of ZRS consisting of nearest-neighbor two holes.
Our systematic exploration provide valuable knowledge on the role of the artificial symmetry breaking on the two-hole GS nature and serves as the
starting point of more sophisticated many-body simulations to uncover more interesting physics of multi-orbital Hubbard model within the symmetry breaking setup. 
\end{abstract}

\maketitle

\section{Introduction}
There have been early theoretical proposals that if the 2$p$ orbitals of nearest-neighbor planar oxygen atom within a unit cell experience the Coulomb repulsion in addition to the conventional onsite Hubbard interaction on $d$ orbitals, then there will be spontaneous splitting of their energy levels~\cite{kivelson2004quasi,fischer2011mean,bulut2013spatially,fischer2014nematic,maier2014pairing,tsuchiizu2018multistage,chiciak2018magnetic,yamase2021theoretical}. 
Intriguingly, this intra-unit-cell orbital ordering leading to the symmetry breaking of the charge transfer energy of the two oxygen atoms has recently been observed experimentally~\cite{epxepy2024}, which found that the $O_x$ and $O_y$ sites within one unit cell exhibit distinct charge transfer energies with separation $\sim 50$ meV.

This fascinating anisotropic  lifting of O-2$p$ energy degeneracy straightforwardly motivates an extended theoretical investigation by generalizing our previous study of multi-orbital Hubbard model in the context of cuprate and nickelate superconductors with impurity approximation~\cite{Mi2020,Mi20,Mi22,Mi23,Mi23H} by imposing additional inter-oxygen repulsion $V_{pp}$. 
In that approach, the full transitional symmetry of the Cu/NiO$_2$ lattice is approximated by an Cu/Ni impurity embedded in the center of the lattice formed by the O ions, whose corresponding band structure are restored with a large finite lattice. Despite that the full lattice's translational invariance is sacrificed, instead of solely taking into account $3d_{x^2-y^2}$ orbital as in the celebrated Emery model~\cite{emery}, this method includes the full multiplet structure of all five $3d$ orbitals of the local Cu/Ni impurity~\cite{zaanen1987electronic,eskes1988tendency,eskes1990cluster}. There are a variety of experimental evidence on the multiplet effects, supporting the triplet states arising from the non-planar orbitals due to Hund's rule, e.g. Auger spectroscopy~\cite{sawatzky1980auger} and X-ray absorption (XAS)~\cite{feiner1992apical} experiments.

\begin{figure}[h!]
\psfig{figure=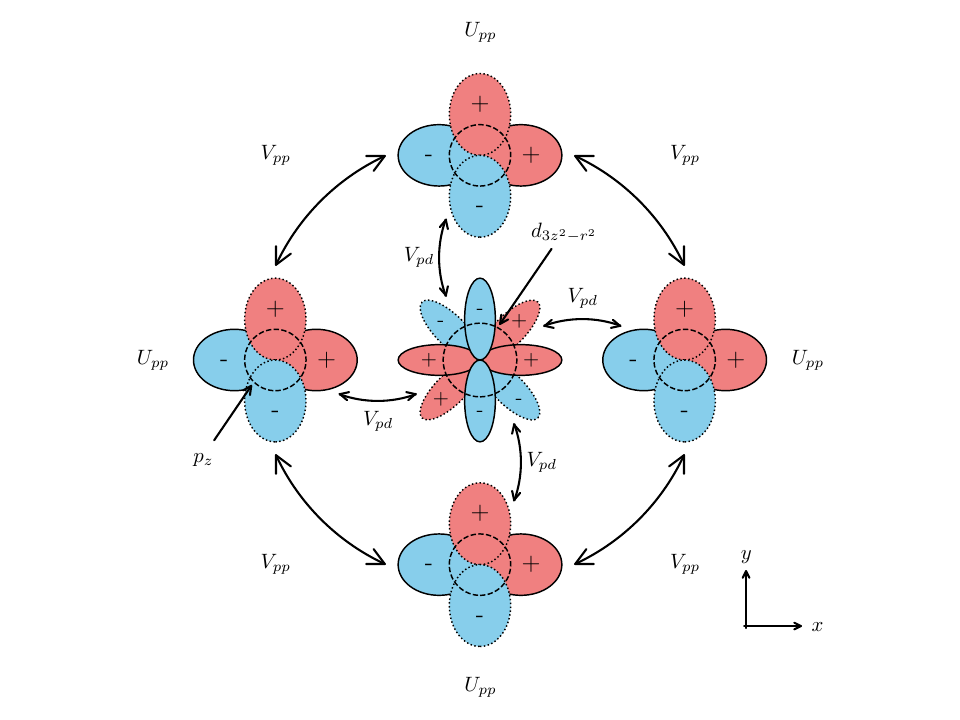,height=8cm,width=.59\textwidth, trim={3cm 1cm 0 0.8cm}, clip} 
\caption{Schematic diagram of CuO$_2$ lattice plane explicitly showing $d_{x^2-y^2}, d_{xy}, p_x, p_y$ orbitals and the intra- and inter-orbital interactions. The black dashed circles denotes $d_{z^2}$ and $p_z$ orbitals while the remaining $d_{xz/yz}$ orbitals are not shown for clarity. The hybridizations between orbitals, e.g. $t_{pd}, t_{pp}$, are not shown for simplicity.}
\label{fig1}
\end{figure}

Because the spontaneous orbital ordering should be essentially a many-body effect, we do not anticipate the observation of this phenomenon by simply adding into the inter-oxygen repulsion $V_{pp}$ in our impurity treatment of the Cu/Ni atom with solely two holes in the whole system~\cite{Mi20,Mi2020,Mi22}. 
Instead, as a complementary study, we will focus on the impact of artificial rotational symmetry breaking by imparting anisotropy to some key parameters such as $d$-$p$ hybridization, onsite energy $\epsilon_p$, the interaction $U_{pp}$ between $O_x$ and $O_y$ orbitals on an O atom, the interaction $V_{pd}$ between NN O and Cu atom.

Undoubtedly, the full many-body treatment of such a multi-orbital model in the thermodynamic limit is impossible nowadays, which is another reason for our present impurity treatment~\cite{qin2022hubbard}. Hence, we follow our previous studies~\cite{Mi20,Mi22} to adopt the single Cu/Ni impurity approximation and particularly focus on the effects of different anisotropy on the local electronic structure, namely the original Zhang-Rice singlet (ZRS) formed by two holes.
%Furthermore, by calculating the Cu/Ni-$3d$ electron removal spectra in various symmetry channels of the $D_{4h}$ point group, we could identify the nature (symmetry, spin, and orbital composition) of the hole-doped states.

%This paper is organized as follows. In Sec.~\ref{model}, we define our model and the variational method employed to study its single-doped hole eigenstates. Sec.~\ref{results} discusses the resulting spectra for various cases considered. The summary and future issues to be addressed are presented in Sec.~\ref{Conclusion}.

%%%%%%%%%%%%%%%%%%%%%%%%%%%%%%%%

\section{Model and Method}\label{model}

To incorporate all essential intra- and inter-orbital interaction effects, the general Hamiltonian reads as follows
\begin{align} \label{H}
	\hat{H} &= \hat{E}^s + \hat{K}^{pd} +\hat{K}^{pp} +\hat{U}^{dd} + \hat{U}^{pp} + \hat{V}^{pp} + \hat{V}^{pd}\nonumber \\
	\hat{E}^s &= \sum_{m\sigma} \epsilon_d(m) \hat{n}^d_{m\sigma}
	+ \sum_{jn\sigma} \epsilon_p(n) \hat{n}^p_{jn\sigma} \nonumber \\ 
	% Kpd
	\hat{K}^{pd} &= \sum_{\langle .j\rangle mn \sigma} 
	(T^{mn}_{pd} \hat{d}^\dagger_{m \sigma}\hat{p}^{\phantom\dagger}_{jn \sigma}+h.c.) \nonumber \\ 
	% Kpp
	\hat{K}^{pp} &= \sum_{\langle jj'\rangle nn' \sigma} 
	(T^{nn'}_{pp} \hat{p}^\dagger_{jn \sigma}\hat{p}^{\phantom\dagger}_{j'n' \sigma}+h.c.) \nonumber \\ 
	% Udd
	\hat{U}^{dd} &= \sum_{\bar{m}_1\bar{m}_2\bar{m}_3\bar{m}_4} U_{dd}(\bar{m}_1\bar{m}_2\bar{m}_3\bar{m}_4) \hat{d}^\dagger_{\bar{m}_1}\hat{d}^{\phantom\dagger}_{\bar{m}_2}\hat{d}^\dagger_{\bar{m}_3}\hat{d}^{\phantom\dagger}_{\bar{m}_4} \nonumber \\
	% Upp
	\hat{U}^{pp} &= \sum_{jnn'\sigma\sigma'} 
	U_{pp}\hat{n}^p_{jn\sigma}\hat{n}^p_{jn'\sigma'}    \nonumber \\
	% Vpp
	\hat{V}^{pp} &= \sum_{\langle jj'\rangle nn'\sigma\sigma'}V_{pp}\hat{n}^p_{jn\sigma}\hat{n}^p_{j'n'\sigma'} \nonumber \\
	%Vpd
    \hat{V}^{pd} &=\sum_{\langle.j\rangle mn\sigma\sigma'}
	V_{pd}\hat{n}^p_{jn\sigma} \hat{n}^d_{m\sigma'}
\end{align}
where $\hat{E}^s$ describes the on-site energies of $3d$ orbital $m$ and $2p$ orbital $n$, in which  ${\hat{n}}_{m\sigma}^d$ is the hole number operator for the $3d$ orbital $m$, whose on-site energy is $\epsilon_d\left(m\right)$. Similarly, $\hat{n}^p_{jn\sigma}$ denotes the hole density in Oxygen's orbital $n\in\{p_x,p_y,p_z\}$ with spin $\sigma$ and at position $j$, whose on-site energy is $\epsilon_p(n)$.

$\hat{K}^{pd}$ and $\hat{K}^{pp}$ describe the Cu-O and O-O hoppings,
respectively. $\langle .j\rangle$ is a sum over the four O adjacent to the Cu impurity and only nearest-neighbor $p-p$ hoppings are retained. Following Slater and Koster\cite{Slater1954}, the Cu-O and O-O hopping integrals
$T^{pd}_{mn}$ and $T^{pp}_{nn'}$ are listed in Table~\ref{table1}. Throughout the paper, all the energies are measured in eV.

In our model we use
$T^{pd}_{b_2}=T^{pd}_{b_1}/2$, so that $t_{pd\pi} = \sqrt{3} t_{pd\sigma}/4$ with $b_1, b_2$ defined in Table~\ref{table1}. We
emphasize that all the Cu-O hybridization parameters $t_{pd}, t_{pp}, t_{pd\sigma}, t_{pd\pi}, t_{pp\sigma}, t_{pp\pi}$ are taken to be positive, and the signs due to the orbitals' overlap are explicitly indicated in Table~\ref{table1} and also depicted in Fig.~\ref{fig1}. The conventional relations $t_{pd} \approx \sqrt{3} t_{pd\sigma}/2 = 2t_{pd\pi}$ and $t_{pp\sigma}=0.9$ eV, $t_{pp\pi}=0.2$ eV are used for our model.

$U_{dd}$ includes the Coulomb and exchange interactions between various $3d^8$ multiplets for all singlet/triplet irreducible representations of the $D_{4h}$ point group in terms of the Racah parameters $A$, $B$ and $C$, which are linear combinations of conventional Slater integrals e.g. $F^k, k=0,2,4$~\cite{pavarini2016quantum}. Here we adopt the shorthand notation $\bar{m}_x \equiv m_x \sigma_x$ where $x=1,\dots,4$ to denote spin-orbitals.
More details on the monolayer situation and the formalism on the calculation of the ground state (GS), precisely the weights of different hole configurations are referred to Ref.~\cite{Mi20,Mi2020,Mi22}. 
Note that precisely the anisotropy lowers the original $D_{4h}$ point group symmetry. However, for simplicity, here we still adopt the interaction matrix for $D_{4h}$ symmetry~\cite{Mi20}, which offer an approximate manner to designate the anisotropy to solely arising from the hopping integrals while retain the lattice symmetry.
Throughout the paper, the conventional values $A=6.0$ eV, $B=0.15$ eV, $C=0.58$ eV are adopted, where $A$ normally characterizes the strength of the magnitude of electron-electron interaction. The Racah parameters $B$ and $C$ are normally set by the atomic physics and not much influenced by screening effects as verified experimentally in other systems with Ni$^{2+}$ ions~\cite{zaanen1990systematics} so that they are conventionally set to be constants~\cite{Mi20,Mi2020,Mi22}. We mention that the conventional Hubbard interaction can be treated as $U_{dd}=A+4B+3C\approx 8.3$ eV in a single-band picture~\cite{Mi20}. 

${\hat{U}}^{pp}$ includes all interaction terms when the two holes are located on the same $O$ site, either the same or different $p$ orbitals. The symbol $U_{pp}$ with subscript denotes the magnitude of ${\hat{U}}^{pp}$. 
In contrast, ${\hat{V}}^{pp}$ represents the nearest-neighbor interaction when two holes are located at nearest-neighbor O-$p$ orbitals with the magnitude $V_{pp}$. 
Besides, ${\hat{V}}^{pd}$ includes the interactions between the impurity Cu-$d$ orbitals and four neighboring O-$p$ orbitals.

Note that our model Eq.~\ref{H} assumes $\epsilon_d=0$ regardless of the specific $d$ orbital for simplicity. This omitting of point-charge crystal splittings is an approximation that the different hybridization with the O orbitals, included in our model, accounts for the difference between the effective on-site energies of the $3d$ orbitals.

Figure~\ref{fig1} depicts the schematic geometry of the most significant orbitals considered in our model.
In the present study, we include all 11 orbitals within a CuO$_2$ unit cell, where $m\in\{a_1, b_1, b_2, e_x, e_y\}$ and $n \in
\{p_{x_1},p_{y_1},p_{z_1},p_{x_2},p_{y_2},p_{z_2} \}$, {\em i.e.} for each O we keep all three O-$2p$ orbitals.

% tpd and tpp:
\begin{table}[h]
\footnotesize
\caption{The Cu-O and O-O hopping integrals $T^{pd}_{mn}$ and $T^{pp}_{nn'}$ with $m \in \{b_1(d_{x^2-y^{2}}), a_1(d_{3z^2-r^{2}}), b_2(d_{xy}), e_x(d_{xz}), e_y(d_{yz}) \}$ and $n \in
\{p_{x_1},p_{y_1},p_{z_1},p_{x_2},p_{y_2},p_{z_2} \}$for our model.Below are exact values of fixed parameters with unit of eV throughout the work. Note that $t_{pd\pi}$ is not listed since it always follows the relation $t_{pd\pi} = \frac{\sqrt{3}}{4} \cdot t_{pd\sigma}$.}
\vspace{1em}

\resizebox{245pt}{!}{
\centering
\begin{tabular}{c|c c c c c c} 
 $m$  & $T^{pd}_{mx_1}$ & $T^{pd}_{my_1}$ & $T^{pd}_{mz_1}$ & $T^{pd}_{mx_2}$ & $T^{pd}_{my_2}$ & $T^{pd}_{mz_2}$ \\ [0.5ex] 
 \hline
 $b_1$ & -$\sqrt{3} t_{pd\sigma}/2$ & $0$ & $0$ & $0$ & $\sqrt{3} t_{pd\sigma}/2$ & $0$ \\
 $a_1$ & -$t_{pd\sigma}/2$ & $0$ & $0$ & $0$ & -$t_{pd\sigma}/2$ & $0$ \\ 
 $b_2$ &  $0$ & $t_{pd\pi}$ & $0$ & $t_{pd\pi}$ & $0$ & $0$ \\
 $e_x$ &  $0$ & $0$ & $t_{pd\pi}$ & $0$ & $0$ & $0$ \\
 $e_y$ &  $0$ & $0$ & $0$ & $0$ & $0$ & $t_{pd\pi}$ \\
 \hline\hline
\end{tabular}} \\

\resizebox{245pt}{!}{
\centering
\begin{tabular}{c c c c c} 
\centering
 2$T^{pp}_{x_1 x_2}$ &  2$T^{pp}_{x_1 y_2}$ & 2$T^{pp}_{x_2 y_1}$ & 2$T^{pp}_{y_1 y_2}$ \\ [0.5ex] 
 \hline
 $t_{pp\pi}-t_{pp\sigma}$ & $t_{pp\pi}+t_{pp\sigma}$ & $t_{pp\pi}+t_{pp\sigma}$ & $t_{pp\pi}-t_{pp\sigma}$ \\ 
 \hline\hline
 \end{tabular}}
 
 \resizebox{200pt}{!}{
 \begin{tabular}{c c c c c c c} 
 \centering
$\epsilon_m$   & A & B & C & $t_{pd\sigma}$ &  $t_{pp\sigma}$ &$t_{pp\pi}$\\ [0.5ex] 
\hline
 0  &  6.0 &  0.15&  0.58 &  1.5 &  0.9 &  0.2 \\
\hline\hline
\end{tabular}}
\label{table1}
\end{table}

Because of the single Cu impurity, we consider the system consisting of only two holes to account for the physics of one doped hole into a $d^9p^6$ system. Hence, the Hilbert space does not grow rapidly with the lattice size, allowing for computational simulations over a large O lattice. Most of the simulations are performed on a  $16 \times 16$ or $20 \times 20$ lattice. We found that the lattice size has negligible effects on the accuracy of the simulations, which essentially arises from the nature of the impurity approximation such that the O sites far away from the central Cu impurity play a minor role.  

With the Hilbert space of all variational states hybridized with various hopping integrals $T^{pd}_{mn}$, $T^{pp}_{nn'}$, and interaction between $3d^8$ multiplets via ${\hat{U}}^{dd}$ as well as ${\hat{U}}^{pp}$, ${\hat{V}}^{pp}$, ${\hat{V}}^{pd}$, we explore the nature of the two-hole ground state (GS) using Exact Diagonalization (ED), specifically focusing on the weights of different two-hole configurations. For example,
\begin{align} \label{eq:HM}
  |GS \rangle = &\sqrt{w_1} |d_{x^2-y^2}d_{x^2-y^2} \rangle  \nonumber\\ + &
  \sqrt{w_2} |d_{x^2-y^2}L_x \rangle \nonumber\\
  + & 
  \sqrt{w_3} |d_{x^2-y^2}L_y \rangle + \dots 
\end{align}
denotes that the GS weights are $w_1, w_2, w_3$ for the corresponding two-hole states. Note that the weights are generically anisotropic, for example, $d_{x^2-y^2}L_x$ and $d_{x^2-y^2}L_y$ can be different.
 
We employ two distinct approaches to introduce anisotropy into the parameters. The first method involves fixing the parameters along the $x$-axis while adjusting those along the $y$-axis to be proportional to the values along the $x$-axis characterized by their ratio $r = t_{pd{\sigma}y}/t_{pd{\sigma}x}$ so that $r = 1$ is the isotropic limit. This approach is exemplified in the discussion of anisotropic $t_{pd}$. The second method maintains a fixed average value for the parameters while ensures that the deviations from this mean along the $x$ and $y$ directions have equal magnitudes but opposite signs. Without loss of generality, in this method, the parameter values along the $x$-axis are set to be larger than those along the $y$-axis. This approach is applied in the discussion of other parameters except $d$-$p$ hybridization.

%%%%%%%%%%%%%%%%%%%%%%%%%%%%%%%%%%%%%%%%%%%%%%%%%%%%%%%%%%%%%%%%%%%%%%%%%%%
\section{Results}\label{results}
%%%%%%%%%%%%%%%%%%%%%%%%%%%%%%%%%%%%%%%%%%%%%%%

As mentioned earlier, the spontaneous orbital ordering due to the additional the Coulomb repulsion between nearest-neighbor planar oxygen atom within a unit cell should be a many-body effect so that our two-hole simulations cannot capture this physics.
To be complete, in Sec.A we first provide our numerical evidence on this expectation. 
Then in Secs.B-D, we focus on the situations of artificially breaking some key parameters e.g. $\epsilon_p$, $t_{pd}$, $U_{pp}$, $V_{pd}$ to have a full understanding of the impact of various anisotropy.

%%%%%%%%%%%%%%%%%%%%%%%%%%%%%%
\subsection{Impact of $V_{pp}$}

\begin{figure} 
\psfig{figure=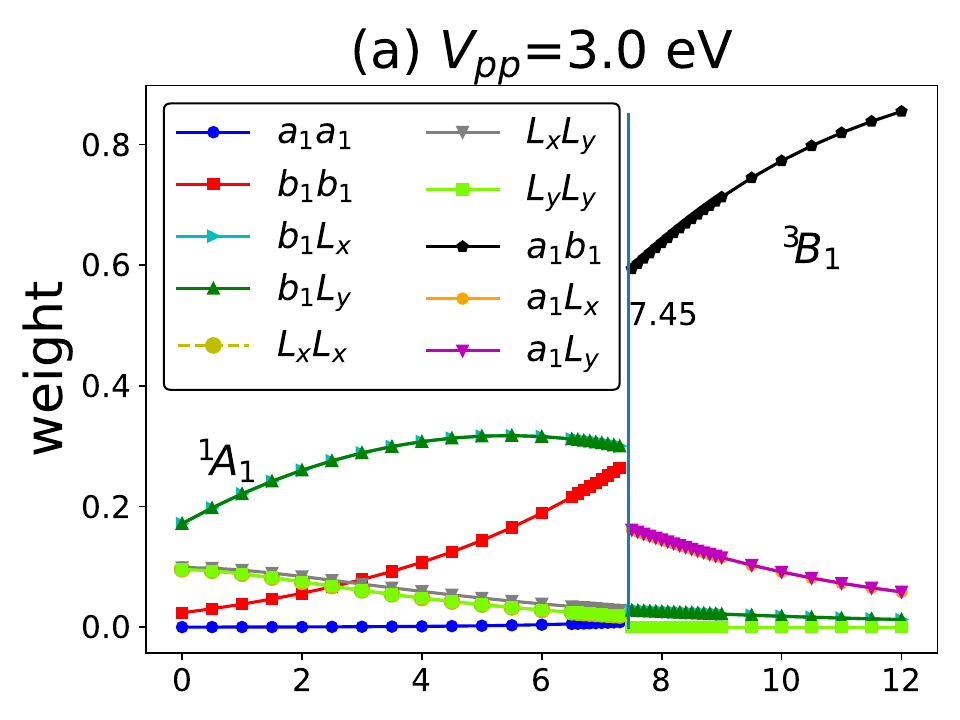,height=5.8cm,width=.49\textwidth, clip} 
\psfig{figure=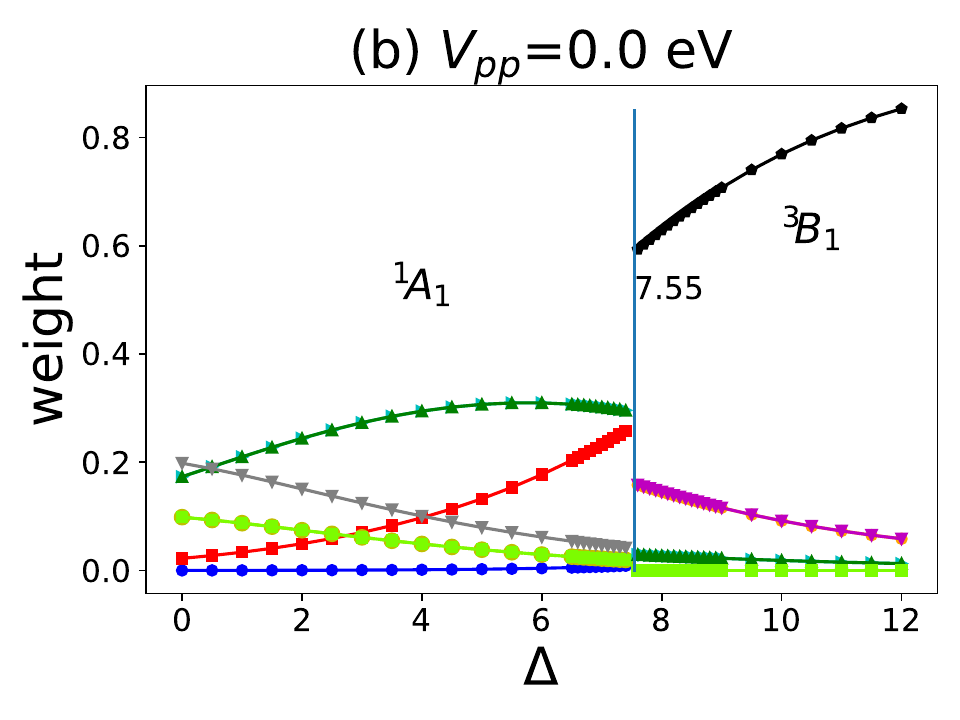,height=5.8cm,width=.49\textwidth, clip} 
\caption{GS weight distribution versus the isotropic charge transfer energy $\Delta=\epsilon_p$ for inter-Oxygen $V_{pp}=3.0, 0.0$ in upper and lower panels separately. $U_{pp}= V_{pd}=0$ and $r=1$ are adopted and other parameters are listed in Table~\ref{table1}. Simply adding into $V_{pp}$ in our model does not induce spontaneous asymmetry between $\hat{x}$ and $\hat{y}$ directions.}
\label{Vpp}
\end{figure}

Figure~\ref{Vpp}'s upper panel illustrates the impact of $V_{pp}$ with a sufficiently large value without anisotropy of any parameters. 
Clearly, it can be seen that even this large $V_{pp}$ does not induce any spontaneous asymmetry between the weights along x and y directions, for instance, the weights of $b_1L_x$ (cyan) and $b_1L_y$ (green) states are exactly the same. 
However, compared with the reference $V_{pp}=0.0$ situation shown in the lower panel, the critical point separating the nature of GS between $^1A_1$ (essentially Zhang-Rice singlet (ZRS)) and $^3B_1$ triplet shifts from 7.55 to 7.45, indicating that the system is more prone to form the $a_1b_1$ triplet state. This is related to the fact that the ZRS state involves one hole on the ligand state that is a linear combination of 4 nearest-neighbor O sites so that the imposed additional $V_{pp}$ tends to avoid this configuration.

Additionally, it is noteworthy that considering $V_{pp}$ results in a significant decrease of the $ L_xL_y $ weight when $ \epsilon_p $ is small. This is related to the repulsive nature of $ V_{pp} $ and also implies for a higher weight of the $ b_1L_{x/y} $ ZRS that is conducive to SC.

%%%%%%%%%%%%%%%%%%%%
\subsection{Anisotropic $\epsilon_p$}

\begin{figure} 
\psfig{figure=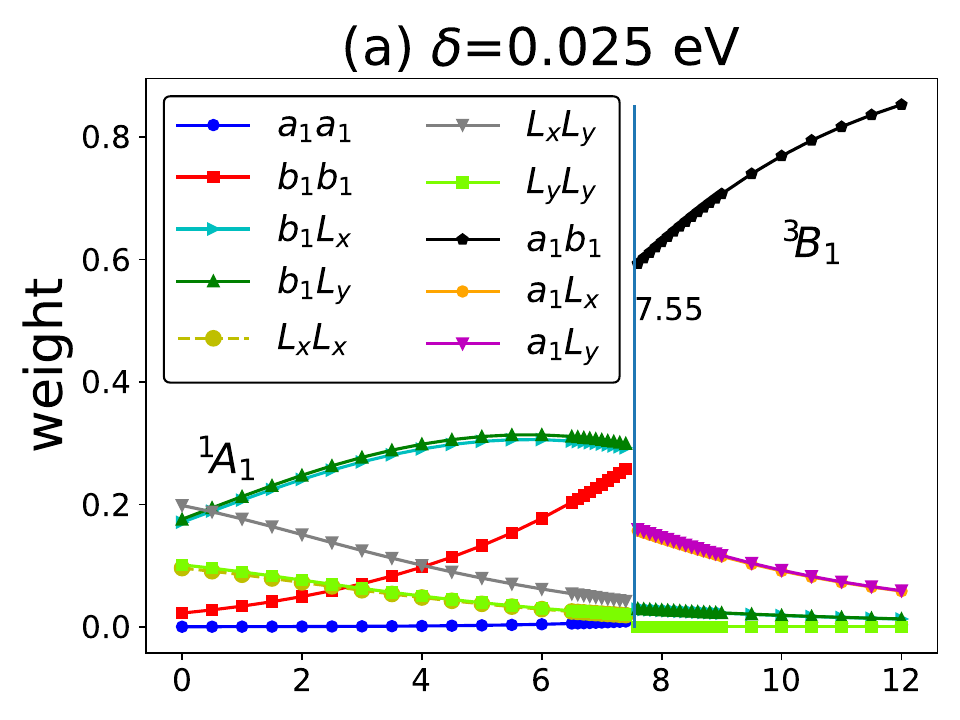,height=5.8cm,width=.49\textwidth, clip} 
\psfig{figure=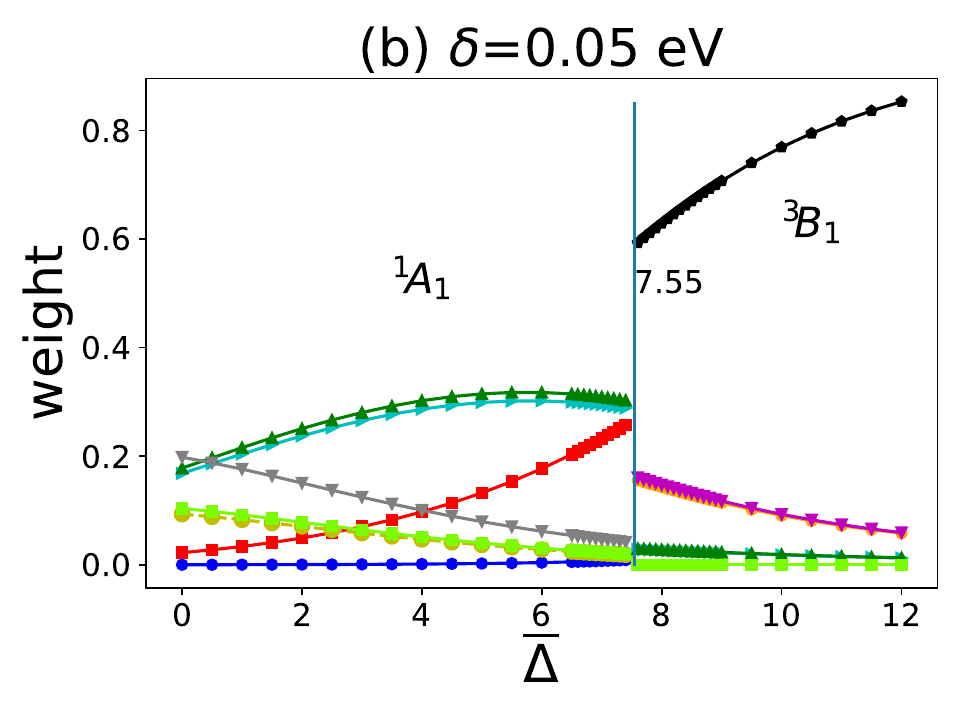,height=5.8cm,width=.49\textwidth, clip}
\caption{GS weight distribution versus average $\bar{\Delta}$  with the variance $\delta=0.025 ,0.05$ in upper and lower panels separately. $V_{pp}$, $U_{pp}$ and $V_{pd}$ are set to zero and $r = 1$. Other parameters are listed in Table~\ref{table1}. 
}
\label{ep}
\end{figure}

\begin{figure}
\psfig{figure=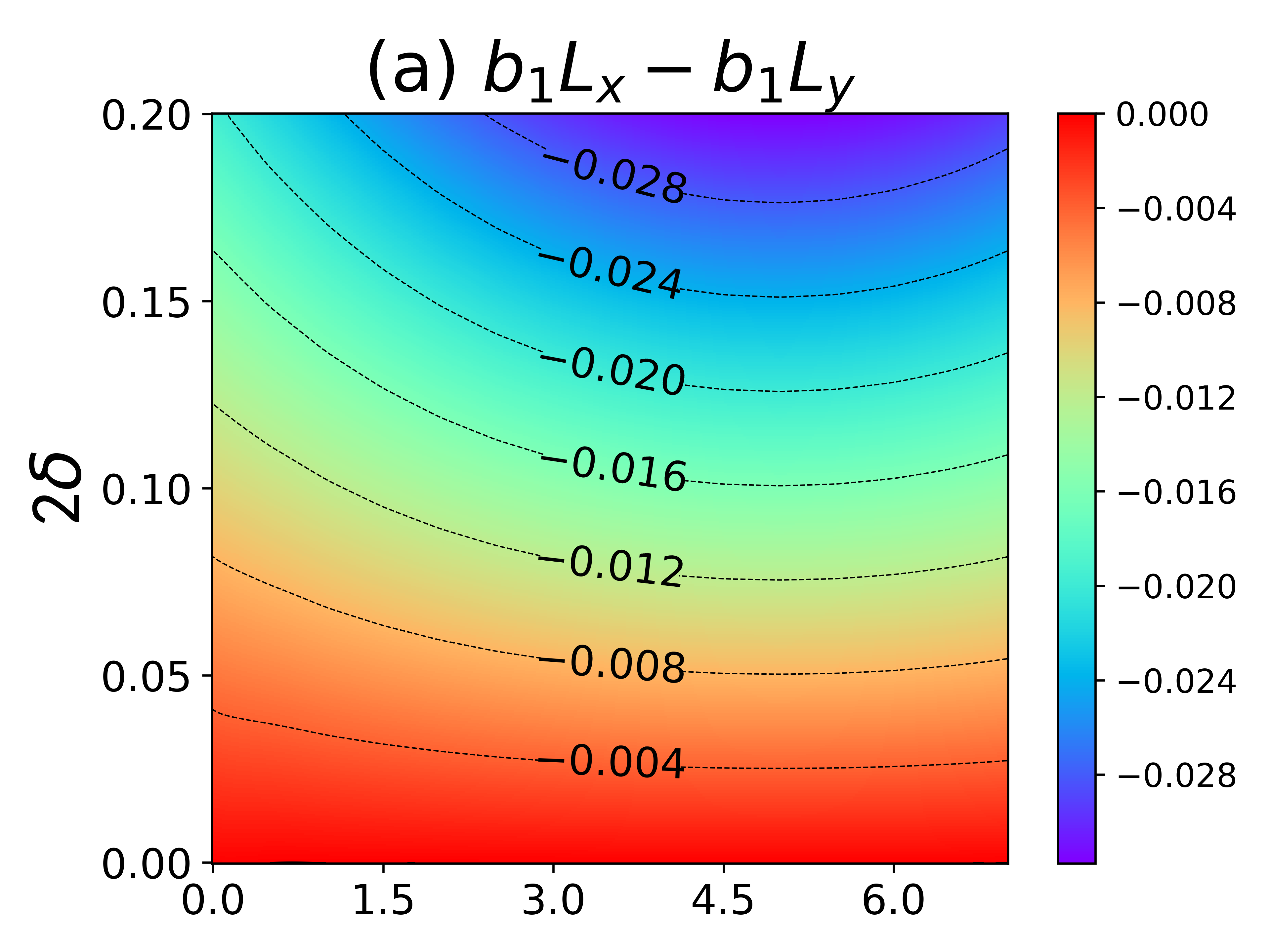,height=5.8cm,width=.49\textwidth, clip} 
\psfig{figure=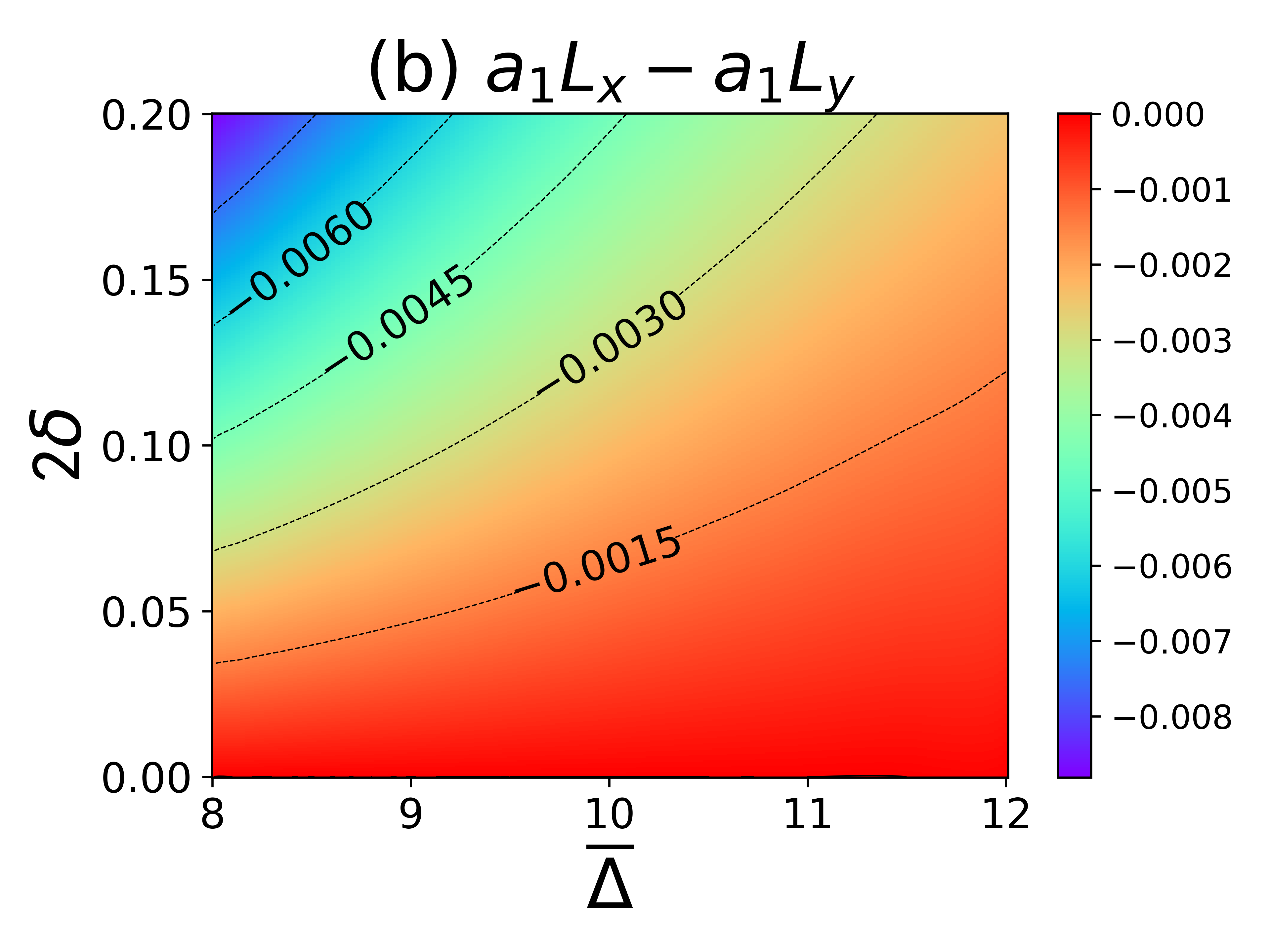,height=5.8cm,width=.49\textwidth, clip} 
\caption{GS weight difference of $b_1L$ or $a_1L$ between $\hat{x}$ and $\hat{y}$ directions influenced by $\bar{\Delta}$ and $2\delta = \epsilon_{px} - \epsilon_{py}$ in upper or lower panels separately. $V_{pp}$, $U_{pp}$ and $V_{pd}$ are set to zero and $r = 1$. Other parameters are listed in Table~\ref{table1}. Since the parameter value in the $\hat{x}$-direction exceeds that in the $\hat{y}$-direction, the weight difference is always negative.
}
\label{ep3d}
\end{figure}

Motivated by the experimental observation of a charge transfer energy splitting of approximately 50 meV~\cite{epxepy2024}, in this section, we set the mean value of $\epsilon_p$ along x and y directions to be $\bar{\Delta}=\bar{\epsilon_p}$ and their difference to be $\delta$ so that the actual values are $\epsilon_{px} = \bar{\Delta} + \delta$ and $\epsilon_{py} = \bar{\Delta} - \delta$. Therefore, the $\delta$ value consistent with the experimental results is 0.025 eV.

Fig.~\ref{ep}(a) shows the evolution of the GS weight versus $\bar{\Delta}$ for the experimentally motivated $\delta=0.025$ eV, which only induces tiny anisotropy. For example, the weight difference for the  ($b_1L_x$, $b_1L_y$), ($L_xL_x$, $L_yL_y$), ($a_1L_x$, $a_1L_y$) state pairs are extremely small. Only at relatively larger $\delta=0.05$ eV can we see the obvious deviation between two directions. Besides, the weights of various states are close to the isotropic system shown in Fig.~\ref{Vpp}'s lower panel and studied before~\cite{Mi20,Mi2020}. 
In addition, similar to the isotropic system, there is a critical $\bar{\Delta}$ separating the two regimes with ($b_1L_x$, $b_1L_y$) dominant ZRS and $a_1b_1$ state dominant respectively. It should not surprising that the critical $\bar{\Delta}$ is independent on the additional asymmetry charge transfer energy between x and y directions, at least in our small $\delta$ regime, since the the average $\delta$ remains unchanged compared with the isotropic system.

Generically, the asymmetry of ($b_1L_x$, $b_1L_y$) 
is slightly larger than ($a_1L_x$, $a_1L_y$). The detailed magnitude of the deviation is mapped in Fig.~\ref{ep3d} in an artificially wide range of $\delta=0-0.2$ eV. The colorbar represents the scale and the vertical axis labels the actual charge transfer energy difference $\epsilon_{px}-\epsilon_{py}=2\delta$ so that remains non-negative by our convention of $\epsilon_{px} \ge \epsilon_{py}$. 
The upper (lower) panel illustrates for the weight difference between $b_1L_x$ and  $b_1L_y$ ($a_1L_x$ and  $a_1L_y$) for small (large) $\bar{\Delta}$ corresponding to the $^1A_1$ ($^3B_1$) GS.

For the upper panel, it can be seen that the asymmetric $\delta$ has stronger impact around the intermediate $\bar{\Delta}\sim 4-5$ eV, where the weight asymmetry between $b_1L_x$ and $b_1L_y$ apparently increases with $\delta$. 
As the average charge transfer energy continues to increase towards the phase transition point around $\bar{\Delta}\sim 7.5$ eV, this influence gradually weakens. 
Conversely, in the lower panel, the weight asymmetry between $a_1L_x$ and $a_1L_y$ is only visible near the phase transition point and then quickly diminishes with further increasing the average charge transfer energy $\bar{\Delta}$.
Moreover, the magnitude of the difference is much smaller than that in the upper panel. This can be naturally understood as the fact that beyond the transition the GS is dominantly $a_1b_1$ state so that the asymmetry reflected by the different charge transfer energy will not be as large as the smaller $\bar{\Delta}$ regime, where the ZRS states lead the physics.

%%%%%%%%%%%%%%%%%%%%%%%%%%%%%%%%%%%
\subsection{Anisotropic $d$-$p$ hybridization}

\begin{figure} 
\psfig{figure=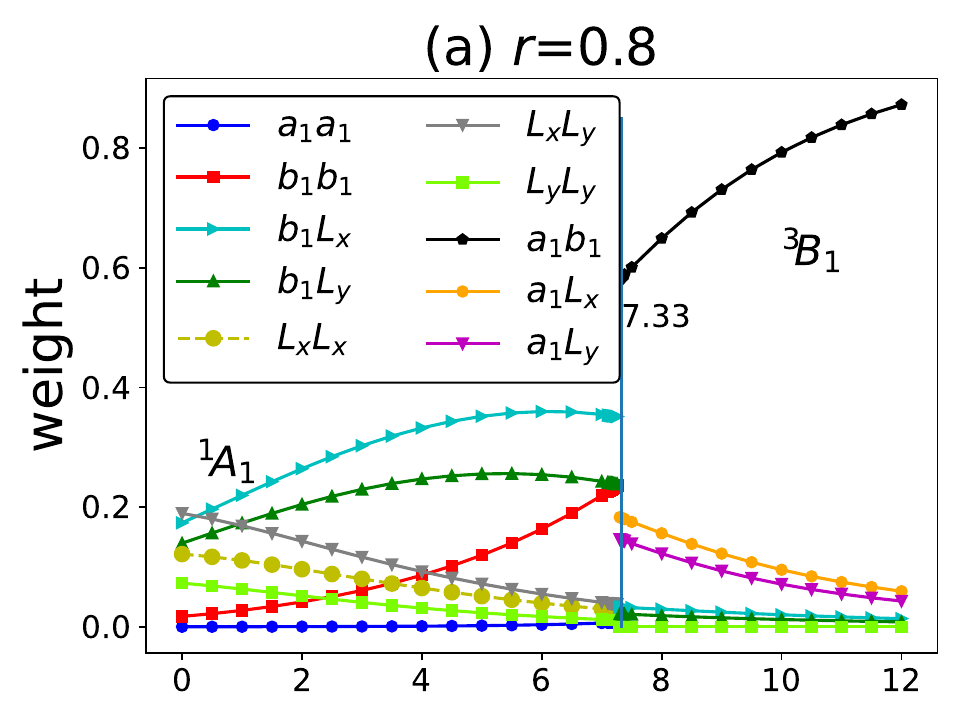,height=5.8cm,width=.49\textwidth, clip} 
\psfig{figure=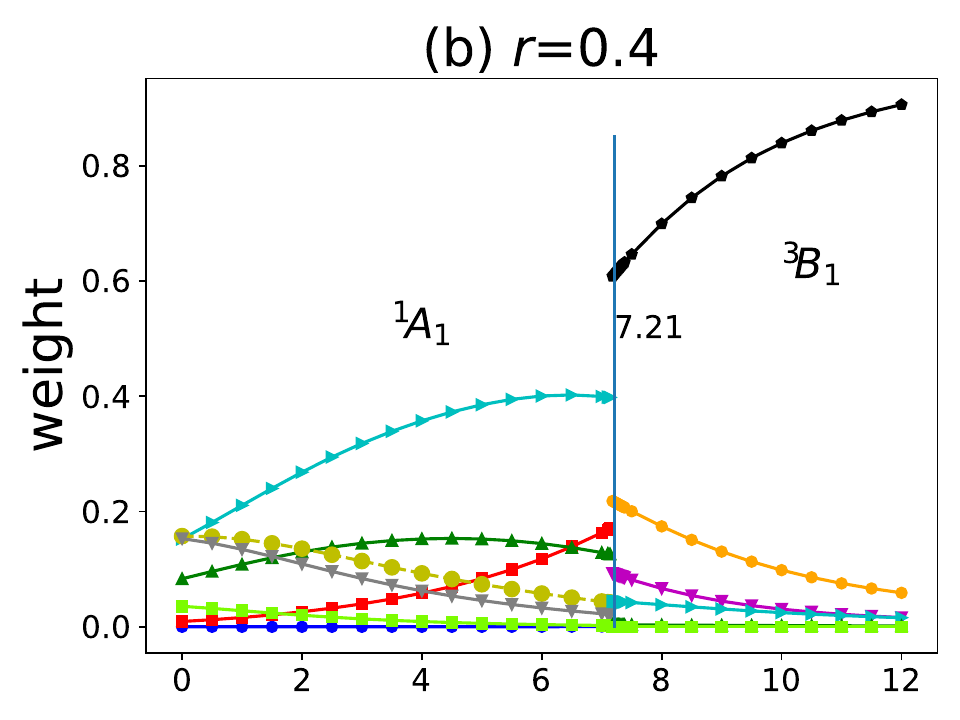,height=5.8cm,width=.49\textwidth, clip} 
\psfig{figure=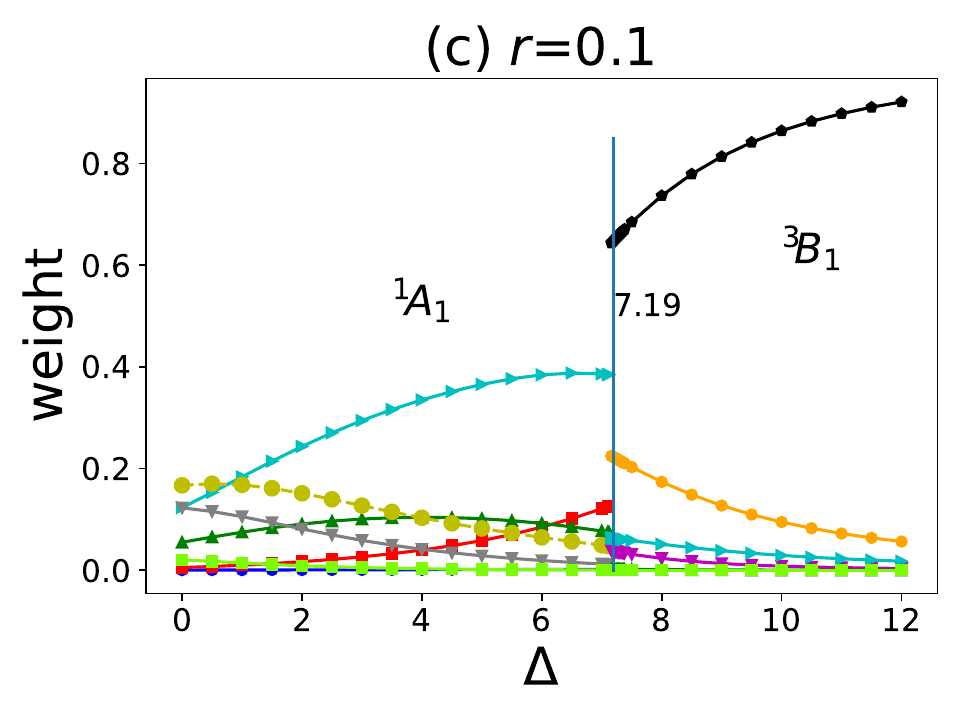,height=5.8cm,width=.49\textwidth, clip} 
\caption{GS weight distribution versus isotropic $\Delta$ for $r = 0.8, 0.4, 0.1$ in upper, middle, and lower panels separately at fixed $t_{pd{\sigma}x}$ = 1.5. $V_{pp}$, $ U_{pp} $, $ V_{pd} $ and $\delta$ are set to zero. Other parameters are listed in Table~\ref{table1}. }
\label{tpd2}
\end{figure}

From now on, we extend the anisotropy to other parameters, in which the most essential one is the $d$-$p$ hybridization.
For $t_{pd}$, we will adopt the first approach to realize its anisotropy, namely by fixing $t_{pdx}$ along $\hat{x}$ direction and vary the ratio $r \equiv t_{pdy}/t_{pdx}$ such that $r=1$ is the isotropic limit.

Fig.~\ref{tpd2} presents the weight distribution for three characteristic $r=0.8, 0.4, 0.1$ from top to bottom panels. The symmetry breaking between x and y directions is significantly reflected by the difference between ($b_1L_x$, $b_1L_y$), ($L_xL_x$, $L_yL_y$), ($a_1L_x$, $a_1L_y$) state pairs while the weight of $L_xL_y$ is not suppressed compared to the isotropic $r=1$ case. 
Undoubtedly, the weight difference boosts with the asymmetry by lowering the ratio. Interestingly, this is realized by the decreasing weight of $b_1L_y$ and $a_1L_y$ while the weight of $b_1L_x$ and $a_1L_x$ almost keep constant, which originates from the unchanged $t_{pd}$ along $\hat{x}$ direction.
Besides, the shift of the phase transition point is more pronounced compared to the impact $V_{pp}$ and anisotropic $\epsilon_p$. We will discuss the critical $\bar{\Delta}$ shift in detail in Fig.~\ref{ratiocp}.

%The last three plots depict the variations of the density of states for the three possibilities of $L^2$: $L_xL_x$, $L_yL_y$, and $L_xL_y$, with respect to both the ratio and $\epsilon_p$. Except for the increasing density of states of $L_xL_x$ with decreasing ratio, the densities of the other two cases decrease. Moreover, no anomalous maximum points similar to those in $b_1L_x$ and $a_1L_x$ are observed, indicating a decrease in the overall Zhang-Rice singlet density when the ratio is small. This reduction is typically unfavorable for superconductivity, suggesting that in this impurity model, the near-one-dimensional limit of the copper oxygen atomic chain is detrimental to the material's superconductivity.

\begin{figure} 
\psfig{figure=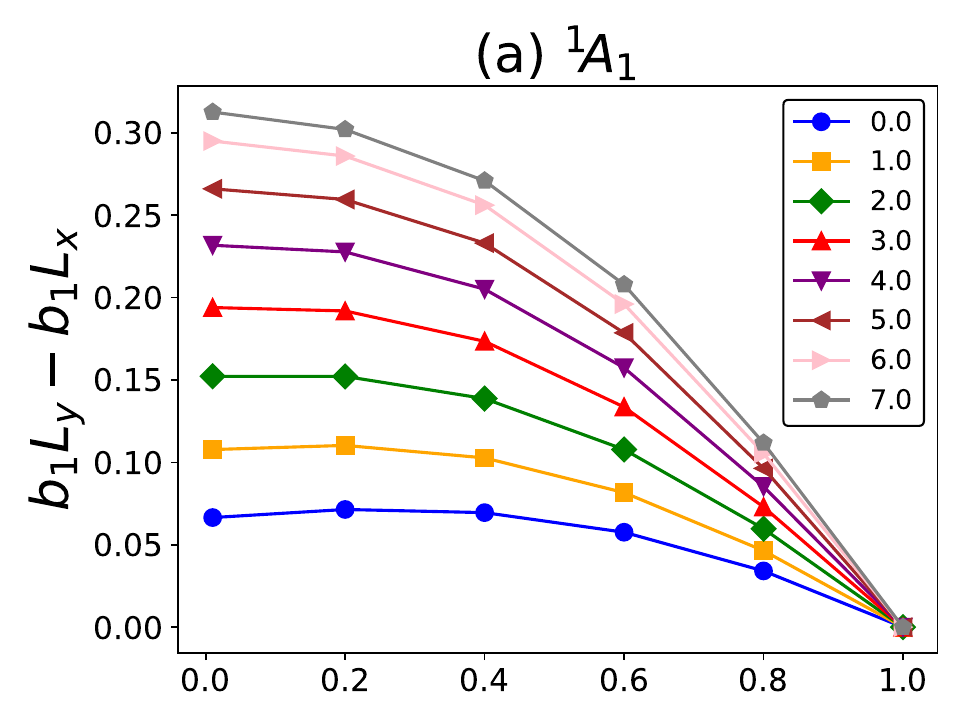,height=5.8cm,width=.49\textwidth, clip} 
\psfig{figure=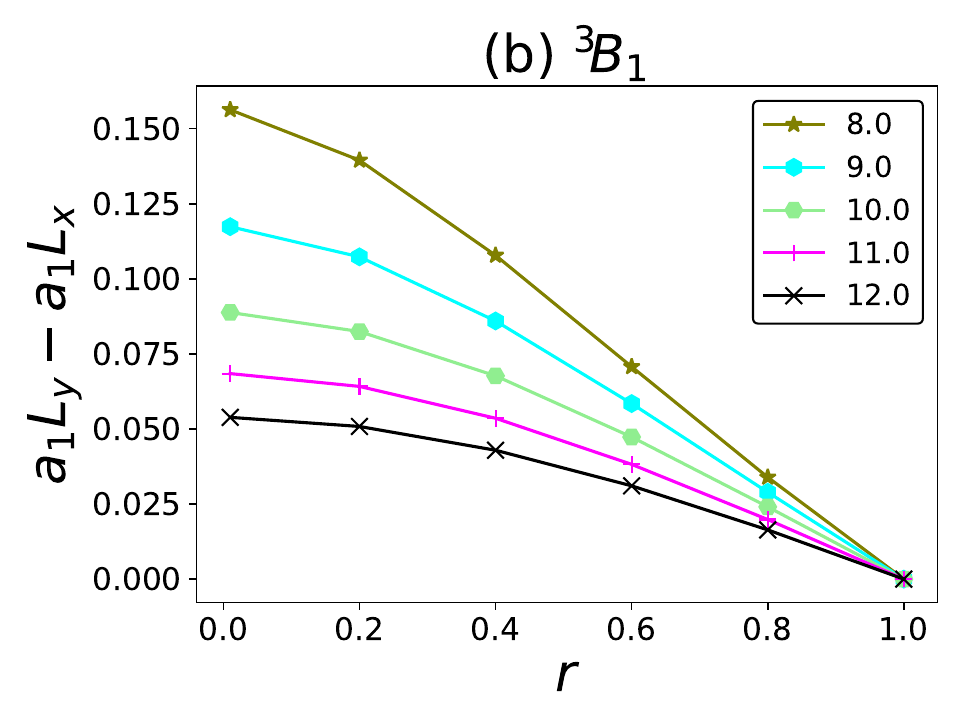,height=5.8cm,width=.49\textwidth, clip} 
\caption{GS weight difference of $b_1L$ or $a_1L$ between $\hat{x}$ and $\hat{y}$ directions versus $r$. The upper and lower panels correspond to the GS of singlet $^1A_1$ and triplet $^3B_1$ nature separately. $V_{pp}$, $ U_{pp} $, $ V_{pd} $ and $\delta$ are set to zero. Other parameters are listed in Table~\ref{table1}. }
\label{tpd3}
\end{figure}

More detailed exploration on the effects of anisotropic $t_{pd}$ follows in order.
Firstly, Fig~.\ref{tpd3} illustrates the most important weight asymmetry of ($b_1L_x$, $b_1L_y$) and ($a_1L_x$, $a_1L_y$) state pairs versus the ratio for various $\Delta$ residing on the two sides of the transition, which is illustrated as the upper and lower panels respectively. Apparently, its evolution with the ratio $r$ is always monotonic, namely smaller ratio so that stronger asymmetry induces larger weight difference. Nonetheless, the dependence on $\Delta$ manifests some non-monotonic effects. Specifically, when $\Delta$ approaches to both small and large limit, the weight difference gradually diminishes. On the one hand, in the small $\Delta$ limit, the two holes can both locate on the O sites, as can be seen from the sizable weight of $L_xL_x$ and $L_yL_y$, so that the dominant state will not only be the conventional ZRS. On the other hand, for large enough $\Delta$, the holes preferentially occupy the $d$ orbitals so that the asymmetric $d$-$p$ hybridization plays a minor role. With these considerations, the weight asymmetry is most pronounced in the intermediate $\Delta \sim 7$ eV regime around the phase transition point. Additionally, the slope of all the curves indicate noticeably steeper feature at large ratios. In other words, when the asymmetry starts to occur, the system has stronger response on the weight redistribution among x and y directions. This rapid weight asymmetry gradually saturates with further breaking the symmetry by lowering the ratio.

It is interesting to ask how the critical $\Delta$ separating the two-hole GS of singlet $^1A_1$ and triplet $^3B_1$ nature reacts to the hybridization ratio $r$.
Fig.~\ref{ratiocp} displays the $r$-$\Delta$ phase diagram with red solid line highlighting the critical charge transfer energy. The generic feature lies in the decreasing critical $\Delta$ with hybridization asymmetry (lowering $r$), which is naturally expected since the symmetry breaking between x and y directions does not prefer the formation of ZRS anymore so that the triplet $^3B_1$ consisting of $a_1b_1$ orbitals is more favorable.
More remarkably, the critical line has a big jump around $\Delta=7.21$, where the ratio spans quite a wide range $r\sim 0.15-0.6$. This implies that the critical behavior of the system is robust against such a large regime of hybridization asymmetry ratio.

\begin{figure}
\psfig{figure=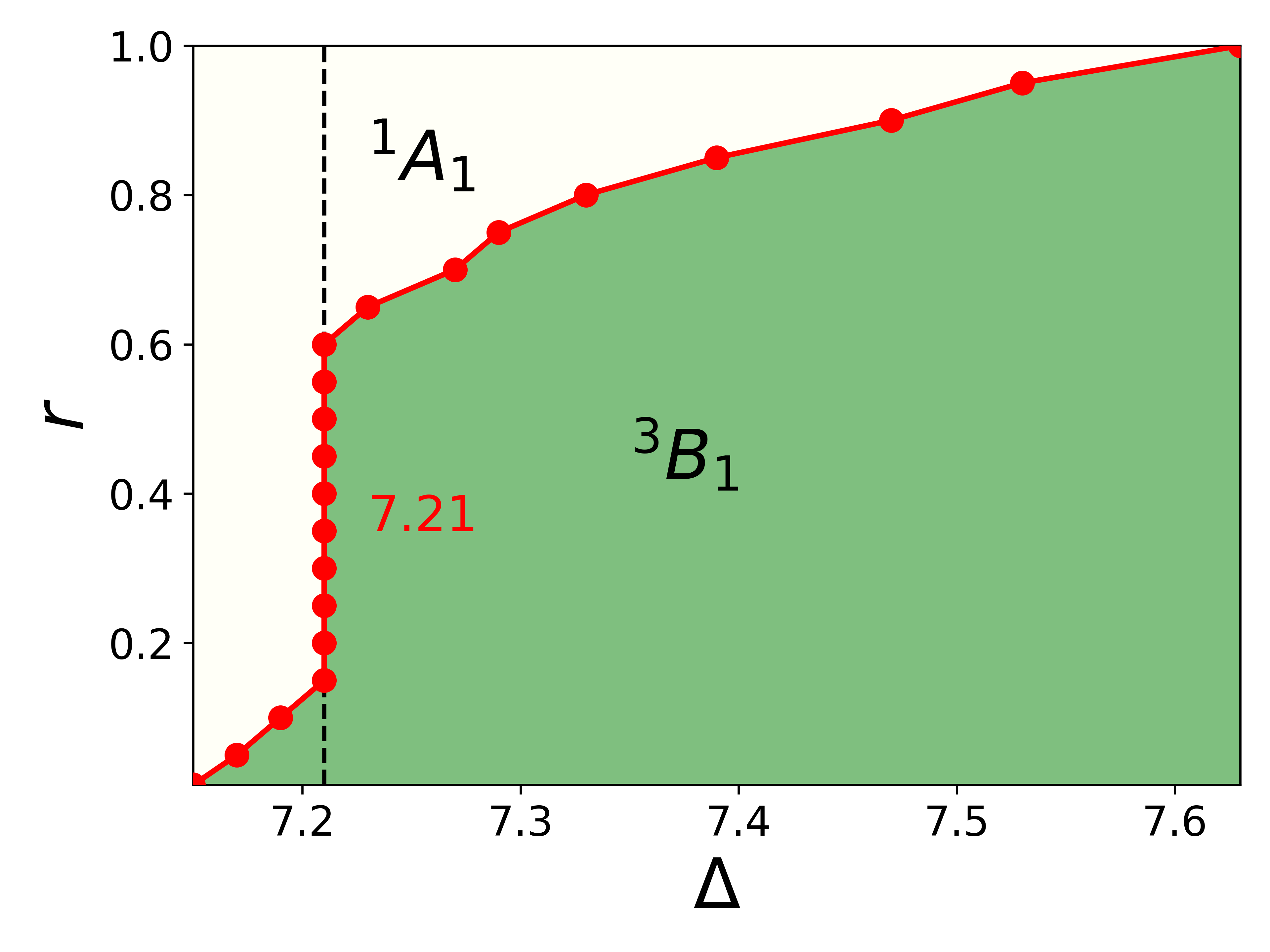,height=5.8cm,width=.49\textwidth, clip} 
\caption{Phase diagram of the ratio $r$ and charge transfer energy $\Delta$. The red solid line represents the critical $\Delta$ separating the GS nature of $^1A_1$ symmetry in the upper left (ivory) and $^3B_1$ symmetry in the lower right (green). $V_{pp}$, $ U_{pp} $, $ V_{pd} $ and $\delta$ are set to zero. Other parameters are listed in Table~\ref{table1}.}
\label{ratiocp}
\end{figure}

%%%%%%%%%%%%%%%%%%%%%%%%%%%%%%%%%%%

\subsection{Anisotropic $U_{pp}$ and $V_{pd}$ interaction}

\begin{figure}
\psfig{figure=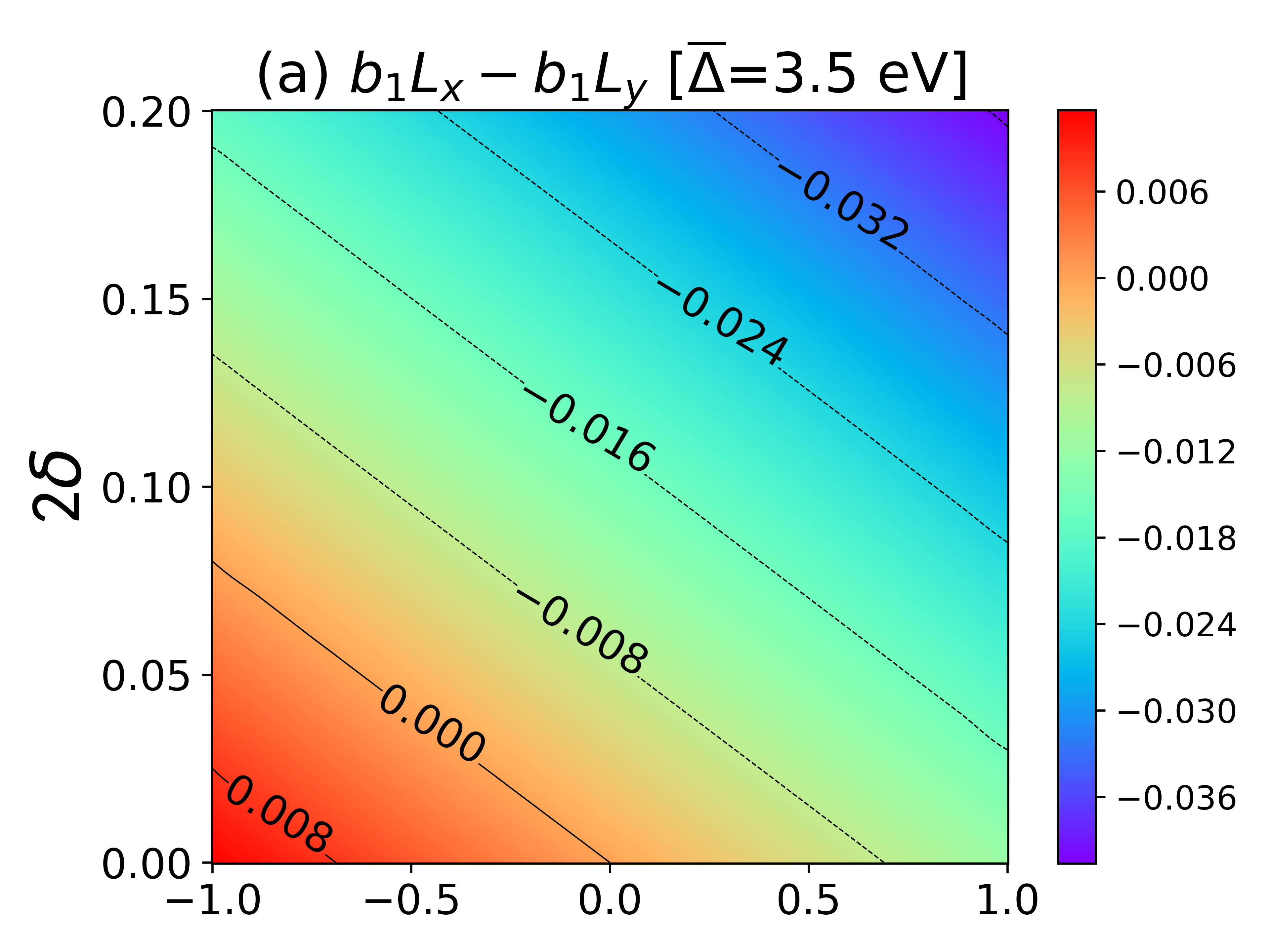,height=5.8cm,width=.49\textwidth, clip} 
\psfig{figure=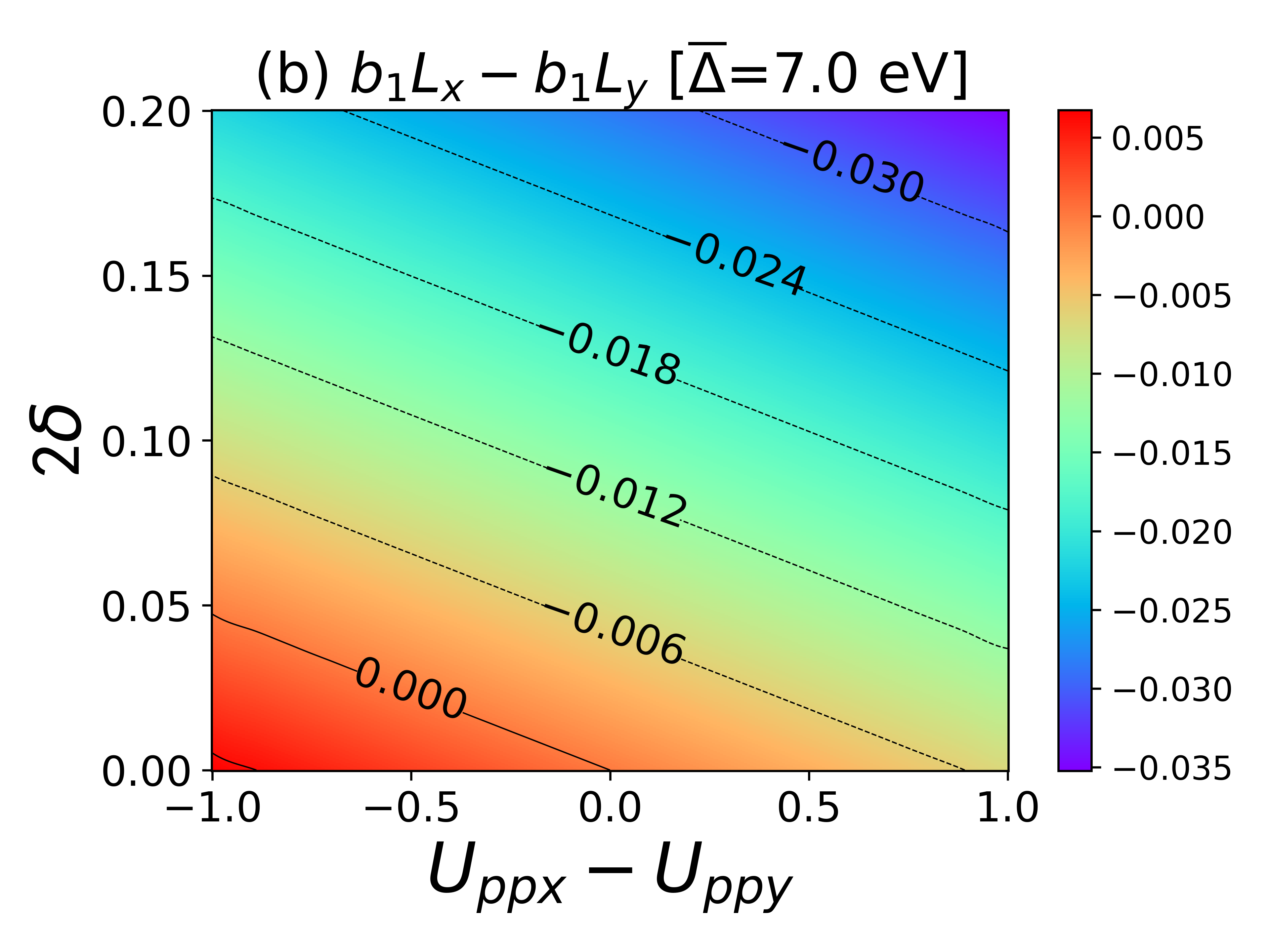,height=5.8cm,width=.49\textwidth, clip} 
\caption{GS weight difference of $b_1L$ between $\hat{x}$ and $\hat{y}$ directions influenced by $U_{ppx}-U_{ppy}$ and  $2\delta$. The upper and lower panels correspond to  $\bar{\Delta} = 3.5,7.0$  with fixed $\bar{U}_{pp}=3.0$. $V_{pp}$ and $ V_{pd} $ are set to zero and $r = 1$. Other parameters are listed in Table~\ref{table1}.}
\label{Upp}
\end{figure}

\begin{figure}
\psfig{figure=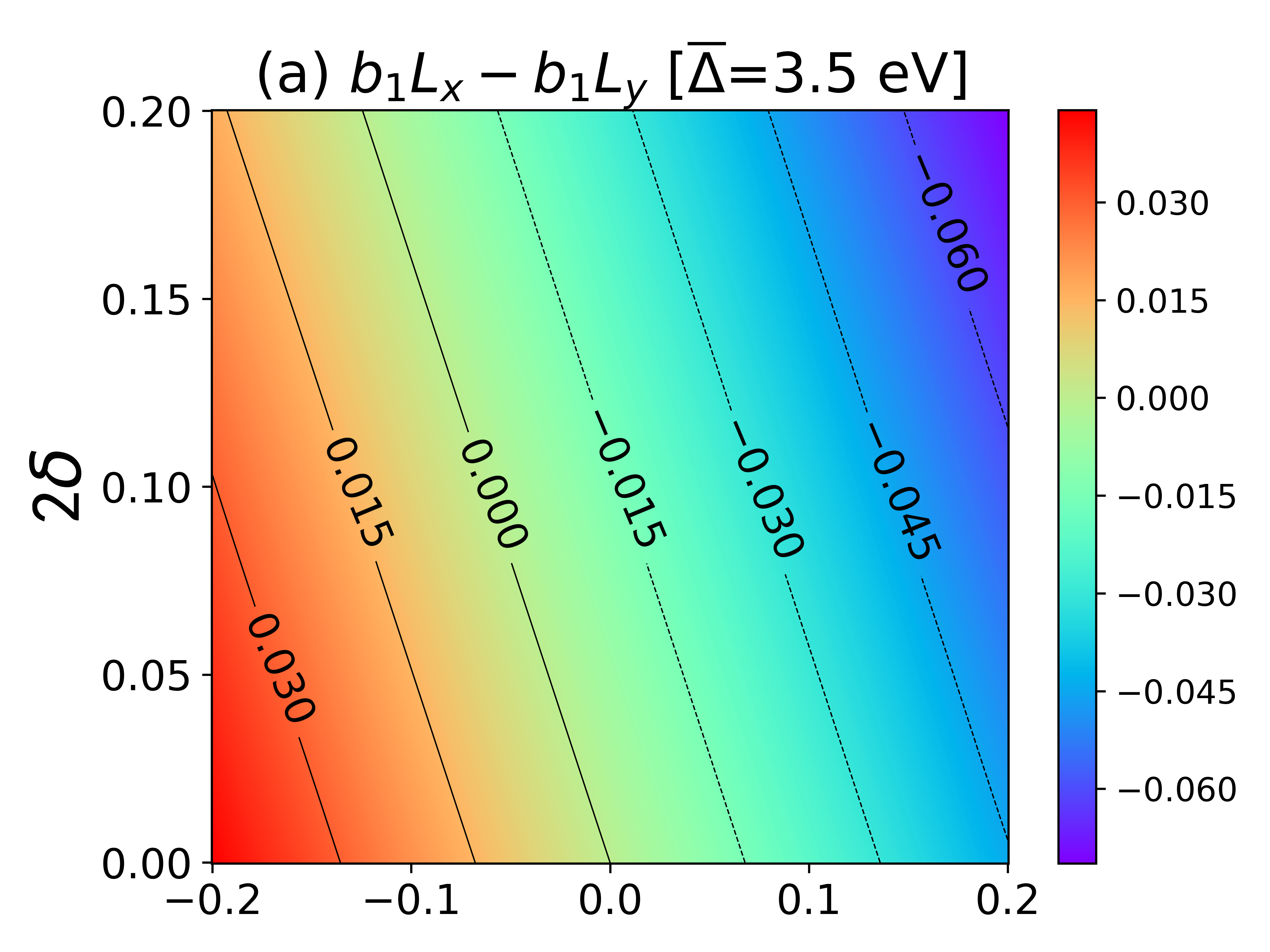,height=5.8cm,width=.49\textwidth, clip} 
\psfig{figure=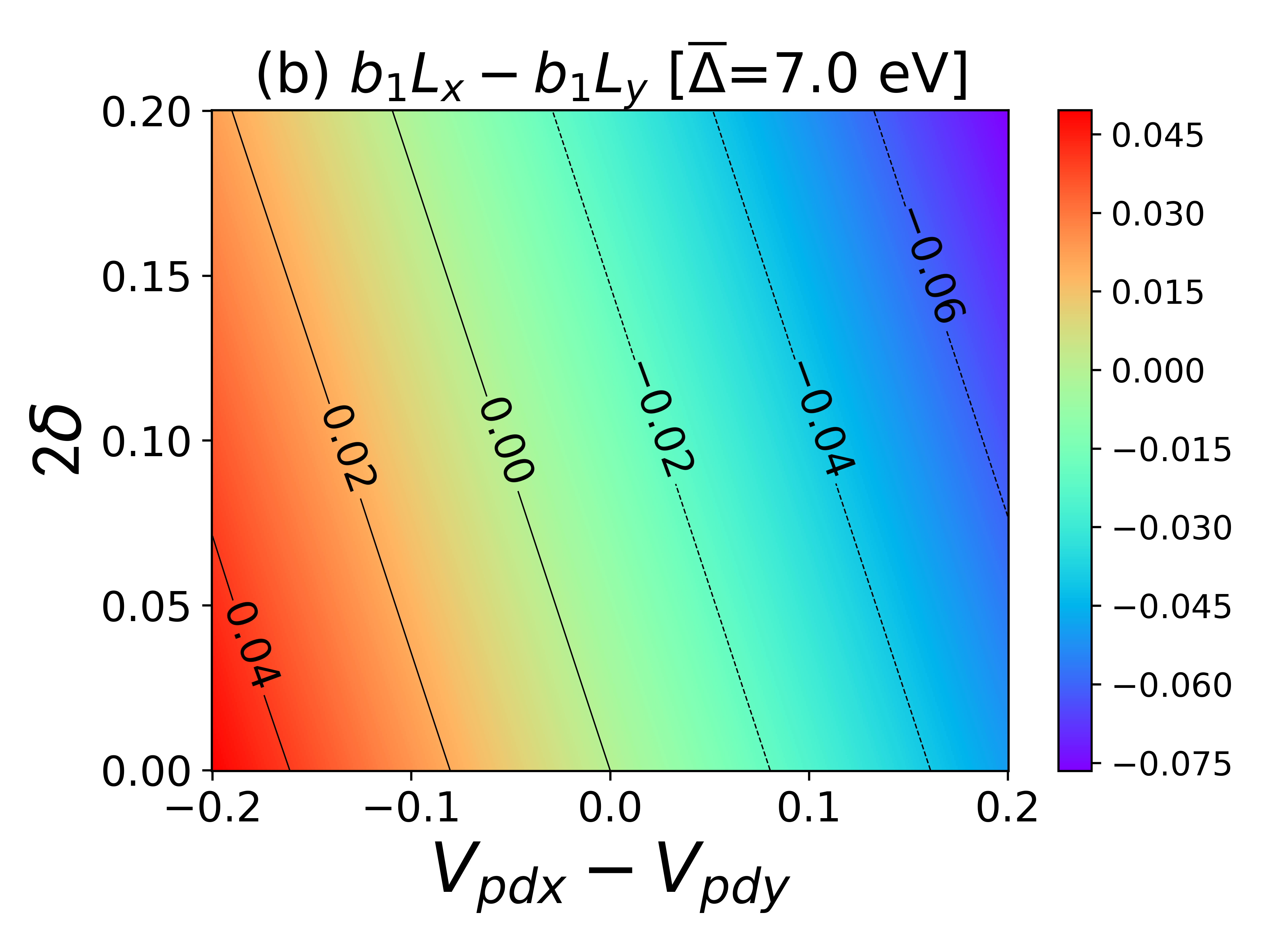,height=5.8cm,width=.49\textwidth, clip} 
\caption{GS weight difference of $b_1L$ between $\hat{x}$ and $\hat{y}$ directions influenced by $V_{pdx}-V_{pdy}$ and  $2\delta$. The upper and lower panels correspond to  $\bar{\Delta} = 3.5,7.0$  with fixed $\bar{V}_{pd} = 0.5$. $V_{pp}$ and $U_{pp}$ are set to zero and $r = 1$. Other parameters are listed in Table~\ref{table1}.}
\label{Vpd}
\end{figure}

Until now we have explored the impact of the asymmetry of two most essential parameters: charge transfer energy $\Delta$ and $d$-$p$ hybridization. In fact, other parameters can also be important in the multi-orbital Hubbard model. One important example is the interaction scale of O ions, namely $U_{pp}$, as well as the nearest-neighbor interaction $V_{pd}$ between $d$-$p$ orbitals. The investigation of the asymmetry of these additional parameters can provide further information on their artificial symmetry breaking and corresponding consequence.
To this aim, Fig.~\ref{Upp} presents the colormap of the weight difference betwen $b_1L_x$ and $b_1L_y$ as a function of the net asymmetric parameter $U_{ppx}-U_{ppy}$ and the net charge transfer energy splitting $\epsilon_{px}-\epsilon_{py}=2\delta$, where $U_{ppx}$ and $U_{ppy}$ denoting the onsite Hubbard interaction of the O sites along x and y directions. 
The upper and lower panels correspond to average $\bar{\Delta} = 3.5$ and $7.0$, which are characteristic values for cuprates and nickelates separately, with fixed average $U_{pp}=3.0$ eV. 

The rich colormap demonstrate quite a few noticeable features. Firstly, the lines in the figure highlight iso-value contours of various weight difference, whose linearity indicates the competition between the local repulsion $U_{ppx}$ disfavoring the double occupancy and small $\epsilon_{px}$ preferring for hole occupation of O sites along $\hat{x}$ direction. 

Secondly, the slope of the line labeled as zero weight difference signals the relative importance between $U_{ppx}$ and $\epsilon_{px}$. Clearly, the upper panel with average $\bar{\Delta} = 3.5$ has a larger slope, which implies that the local repulsion $U_{ppx}$ has more important impact than $\epsilon_{px}$. This can be exemplified when $U_{ppx}-U_{ppy}=-2.0$ eV, where the critical $\epsilon_{px}-\epsilon_{py} \sim 0.075$ for $\bar{\Delta}=3.5$ is larger than the critical $\epsilon_{px}-\epsilon_{py} \sim 0.05$ for $\bar{\Delta}=7.0$. This is simply the consequence of the fact that larger $\bar{\Delta}$ forbids the hole occupancy on O sites so that O's local interaction is not important at all. Therefore, it can be expected that the lines in the colormap will become flatter with further increasing the charge transfer energy $\bar{\Delta}$.

Finally, all the iso-value lines have the same slope, which implies that the relative impact of the asymmetry of charge transfer energy and local repulsion is independent on their specific value regime. 
In fact, this remains true in
Fig.~\ref{Vpd}, where the local repulsion is replaced by the nearest-neighbor $d$-$p$ repulsion $V_{pd}$.
The dominant feature lies in the much larger slope of iso-value lines so that $V_{pd}$ has stronger impact than $U_{pp}$. This is owing to the nature of $V_{pd}$ itself, as its value significantly affects the ZRS, where the two holes would experience this repulsion directly in spite of its small average magnitude around 0.5 eV.
Besides, the comparison between two panels demonstrates that larger $\bar{\Delta}$ at lower panel has slightly larger slope. Additionally, compared with Fig.~\ref{Upp}, the weight difference displays generically larger magnitude even when $V_{pd}$ is much smaller than $U_{pp}$ so that $V_{pd}$ has much stronger impact on the symmetry breaking within the system, which again stems from the nature of ZRS consisting of nearest-neighbor two holes.

%%%%%%%%%%%%%%%%%%%%%%%%%%%%%%%%%%%%%%%%%%%%%%%%%%%%%%%%%%%%%%%%%%%%%%%%%%%
\section{Summary and outlook}\label{Conclusion}

Motivated by the recent experimental findings on the spontaneous symmetry breaking of the charge transfer energy splitting of the two oxygen atoms within a unit cell~\cite{epxepy2024}, we have investigated the multi-orbital Hubbard model adopting the impurity approximation, where the Cu ion is treated as an impurity sitting at the center of O lattice considering all the Cu-3$d^8$ multiplet interaction as well as the local and nearest-neighbor interaction between Cu-3$d$ and O-2$p$ orbitals.

Although the simply additional nearest-neighbor repulsion $V_{pp}$ between O atoms within a unit cell proves to be insufficient to induce the spontaneous orbital ordering in our simplified impurity model, the artificial breaking of various parameters reveal rich phenomena in terms of the weight asymmetry between $\hat{x}$ and $\hat{y}$ directions.
We have mapped out different phase diagram of the weight redistribution versus a few essential parameters, e.g. charge transfer energy, $d$-$p$ hybridization, local repulsion $U_{pp}$, and $d$-$p$ repulsion $V_{pd}$. The numerical evidence reveal that the ZRS can be affected by these parameters to distinct extent. Although the experimentally motivated asymmetric charge transfer energy only induces tiny weight difference, the asymmetric $d$-$p$ hybridization can result in considerable change of the weight, which also remarkably shows that the critical charge transfer energy separating the $^1A_1$ singlet and $^3B_1$ triplet GS is independent on intermediate asymmetry ratio regime. Besides, the nearest-neighbor $V_{pd}$ has much stronger impact than the local $U_{pp}$ on the symmetry breaking within the system, which stems from the nature of ZRS consisting of nearest-neighbor two holes.

Our systematic investigation provide valuable knowledge on the role of the artificial symmetry breaking on the two-hole GS nature. 
Nonetheless, it is worthwhile emphasizing that the current study is only the starting point because of the impurity approximation of the $d$ orbital and only two holes within the whole system does not take into account more complicated many-body effects.
The more synthetic investigation with the advanced many-body numerical methods is requisite and essential to uncover more interesting physics of multi-orbital Hubbard model within the symmetry breaking setup and is in progress.

\section{Acknowledgements} 
This work was supported by National Natural Science Foundation of China (NSFC) Grant No. 12174278, startup fund from Soochow University, and Priority Academic Program Development (PAPD) of Jiangsu Higher Education Institutions.

%%%%%%%%%%%%%%%%%%%%%%%%%%%%%%%%%%%%%%%%%%%%%%%%%%%%%%%%
\bibliographystyle{apsrev4-1}
\bibliography{main}

\end{document}